\documentclass[aps,twocolumn,preprintnumbers,prd,superscriptaddress,10pt,nofootinbib]{revtex4-1}

\usepackage{amssymb,amsmath,amsthm,amsfonts}
\usepackage[usenames]{color}
\definecolor{darkred}{rgb}{0.5,0,0}

\usepackage{bm}
\usepackage{dcolumn}
\usepackage[utf8]{inputenc}
\usepackage{latexsym}
\usepackage{rotating}
\usepackage{longtable}

\usepackage{enumerate}
\usepackage{tensor,multirow}
\usepackage{url}
\usepackage[linktocpage]{hyperref}
\usepackage{float}
\usepackage{graphicx}

\def\be{\begin{equation}}
\def\ee{\end{equation}}
\newcommand{\beq}{\begin{eqnarray}}
\newcommand{\eeq}{\end{eqnarray}}

\def\ba{\begin{align}}
\def\ea{\end{align}}

\begin{document}

\title{Addendum to ``Strong cosmic censorship: The nonlinear story''}

\author{Raimon~Luna}
\affiliation{Physique Th\'{e}orique et Math\'{e}matique, Universit\'{e} libre de Bruxelles and International Solvay Institutes, Campus Plaine C.P. 231, B-1050 Bruxelles, Belgium}

\author{Miguel~Zilh\~ao}
\affiliation{CENTRA, Departamento de F\'{\i}sica, Instituto Superior T\'ecnico
  -- IST, Universidade de Lisboa -- UL, Avenida Rovisco Pais 1, 1049 Lisboa,
  Portugal}

\author{Vitor~Cardoso}
\affiliation{CENTRA, Departamento de F\'{\i}sica, Instituto Superior T\'ecnico
  -- IST, Universidade de Lisboa -- UL, Avenida Rovisco Pais 1, 1049 Lisboa,
  Portugal}

\author{Jo\~ao~L.~Costa}
\affiliation{Departamento de Matem\'atica, ISCTE - Instituto Universit\'ario de
  Lisboa, Portugal}
\affiliation{Center for Mathematical Analysis, Geometry and Dynamical Systems,
  Instituto Superior T\'ecnico -- IST, Universidade de Lisboa -- UL, Avenida
  Rovisco Pais 1, 1049 Lisboa, Portugal}

\author{Jos\'e~Nat\'ario}
\affiliation{Center for Mathematical Analysis, Geometry and Dynamical Systems,
  Instituto Superior T\'ecnico -- IST, Universidade de Lisboa -- UL, Avenida
  Rovisco Pais 1, 1049 Lisboa, Portugal}

\begin{abstract}
We clarify a number of issues that arise when extending the analysis of Strong Cosmic Censorship (SCC) to perturbations of highly charged Reissner-Nordstr\"{o}m de Sitter (RNdS) spacetimes. The linear stability of the Cauchy horizon can be determined from the spectral gap of quasinormal modes, thus giving a clear idea of the ranges of parameters that are likely to lead to SCC violations for infinitesimally small perturbations. However, the situation becomes much more subtle once nonlinear backreaction is taken into account. These subtleties have created a considerable amount of confusion in the literature regarding the conclusions one is able to derive about SCC from the available numerical simulations.  
 Here we present new numerical results concerning charged self-gravitating scalar fields in spherical symmetry, correct some previous claims concerning the neutral case, and argue that the existing numerical codes are insufficient to draw conclusions about the potential failure of SCC for near extremal RNdS black hole spacetimes. 
\end{abstract}

\maketitle
\section{Introduction}

Strong Cosmic Censorship (SCC) conjectures that Cauchy Horizons (CHs) -- the  boundaries  of  the  maximal  evolution of initial data via the Einstein field equations -- are unstable and  give  rise,  upon  perturbation,  to singular boundaries where the Einstein field equations break down. 

It is important to clarify how strong these singular boundaries must become in order to correspond to {\em terminal boundaries} for the validity of the field equations. In this respect, it is well known that the blow up of curvature is not enough to imply neither the breakdown of the field equations~\cite{L2} nor, in fact,  the  destruction  of  macroscopic  observers~\cite{Ori:1991zz}. So, in order to guarantee the formation of a terminal boundary one needs a stronger type of singularity. In the context of the Einstein-scalar-field system in spherical symmetry, the natural candidate is a {\em mass inflation singularity}, where the Misner-Sharp mass diverges. In this scenario, not only the Kretschmann curvature scalar necessarily blows up\footnote{Note that the converse is not necessarily true, since it is possible to construct solutions with bounded Misner-Sharp mass and diverging Kretschmann scalar~\cite{Costa:2017tjc}.}, but moreover the field equations are expected to break down completely~\cite{Dafermos:2012np}.           

In the context of  linear scalar field perturbations, as studied in~\cite{Cardoso:2017soq}, mass inflation (and the breakdown of the field equations) cannot be studied directly, since the spacetime geometry is fixed.  We can however study, as a proxy for this phenomenon, the blow up of the $H^1$ norm of the scalar field perturbation. The numerical results in~\cite{Cardoso:2017soq} suggest that this blow up does not occur for highly charged/near extremal Reissner-Nordstr\"{o}m de Sitter (RNdS) black holes (BHs), i.e. the CHs in these spacetimes are expected to be linearly stable. 
This suggests a potential failure of SCC! But the confirmation of this disturbing suggestion demands an understanding of nonlinear effects.   

The authors of this note recently discussed a nonlinear numerical analysis of the CH of RNdS spacetimes~\cite{Luna:2018jfk}. The main purpose of that study was to confirm whether the aforementioned linear results of Ref.~\cite{Cardoso:2017soq}  would also hold in the full nonlinear setting. More precisely, the authors wished to confirm if nonlinear perturbations of the RNdS BHs identified in~\cite{Cardoso:2017soq} as possessing linearly stable CHs give rise to spacetimes where the Misner-Sharp mass remains bounded.
 
To ascertain the nonlinear stability of the CH we studied the interaction of a neutral scalar field pulse with the BH (as determined by the Einstein-scalar field system with a positive cosmological constant in spherical symmetry), and checked whether this pulse triggered mass inflation in the BH interior. The corresponding simulations seemed to show, fairly convincingly, that mass inflation completely disappeared above a certain BH charge threshold. We thus concluded from this evidence that nonlinear effects were apparently insufficient to preserve SCC. Naturally, we were aware that no irrefutable proof regarding the asymptotic behavior of the Misner-Sharp mass could be obtained from our numerical evolution: any numerical code has a necessarily finite maximal evolution time, and, in principle, mass inflation could always emerge after such a time. Nonetheless, our claim concerning the inexistence of mass inflation was supported by the fact that our evolution showed a nearly constant Misner-Sharp mass well within the instability (blue-shift) region~\cite{Dafermos:2003wr}, i.e., the Misner-Sharp mass remained small while the Kretschmann scalar was already diverging.

In the meantime, we have improved our numerical code by, in particular, extending its scope to charged scalar fields;  we will present some of the corresponding new numerical results here. As part of this effort, we have also revisited the neutral case and gained new insights into the problem.
In view of these, we feel obliged to retract the claim -- made in Ref.~\cite{Luna:2018jfk} -- regarding the construction of numerical solutions with no mass inflation, and discuss some possible sources of confusion that can arise in the interpretation of the nonlinear effects. We will also argue that, despite the aforementioned improvements, the existing numerical codes are insufficient to draw conclusions regarding the potential failure of SCC for near extremal RNdS BH spacetimes.

\section{New numerical results.}

In Ref.~\cite{Luna:2018jfk} we discussed two sets of simulations, where we considered charged BHs of (initial) mass $M_0=1$, cosmological constant $M_0^2\Lambda=0.06$, and two different values for the BH charge, $Q=0.9$ and $Q=1.0068$.
The first one, $Q=0.9$, corresponding to $Q=0.890Q_{\rm max}$, is below the stability threshold $Q_{\rm th}=0.992Q_{\rm max}$ predicted by~\cite{Cardoso:2017soq}, where $Q_{\rm max}$ is the maximal (extremal) charge of the RNdS solutions with parameters $M$ and $\Lambda$. In this case, the CH is linearly unstable, and, as expected, mass inflation occurred in the numerical simulations. The other value of the charge was $Q=1.0068$, corresponding to $Q=0.996Q_{\rm max}$, which, according to the linear results in Ref.~\cite{Cardoso:2017soq}, is expected to lead to violations of SCC, in the sense that small enough perturbations of these solutions are expected to have bounded Misner-Sharp mass up to the CH. Indeed, in the results shown in Ref.~\cite{Luna:2018jfk} no mass inflation was observed.

However, when sending in a scalar field pulse in a nonlinear evolution, accretion by the BH will necessarily increase its mass, which tends to make it less extremal, i.e., less likely to experience mass inflation according to the linear analysis~\cite{Cardoso:2017soq}. This effect is key, since a large enough scalar field pulse will drive the BH outside the region of expected SCC violation (given by $Q_{\rm th} \sim (1-10^{-3}) Q_{\max}$ for neutral scalar fields, and $Q_{\rm th} \sim (1-10^{-5}) Q_{\max}$ for charged scalar fields, where $Q_{\max}$ should be computed using the final BH mass and charge). If this happens, mass inflation will probably reappear, whether we are able to see it during the course of the numerical evolution or not.

%
%
%
\begin{figure}
\includegraphics[width = 0.49\textwidth]{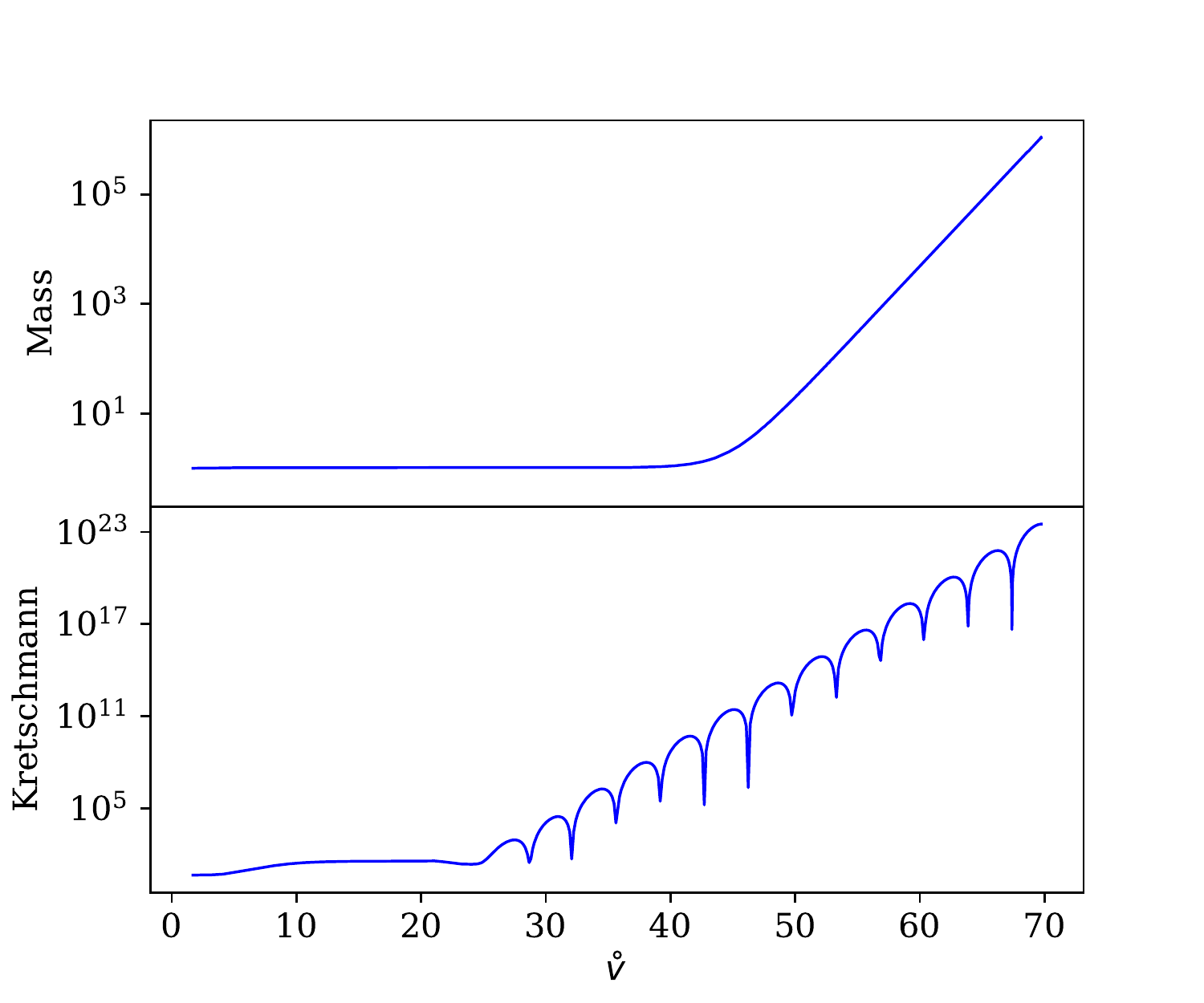}
\caption{Misner-Sharp mass and Kretschmann scalar profiles in the BH interior, as functions of the Eddington-Finkelstein coordinate $\mathring{v}$ along outgoing null geodesics, for a charged ($q = 0.45$) massless scalar field. Parameters: $\Lambda = 0.06$, $Q_0/Q_\text{max} = 1 - 10^{-5}$.\label{Charged}}
\end{figure}
Accretion effects were noted in Ref.~\cite{zhang2019strong}, where it was argued that they would make SCC violations impossible. We cannot agree with this argument: as a matter of principle, one can always use arbitrarily small perturbations, sufficiently small as to keep the BH in the near extremal regime where potential failures of SCC are expected. 

For the simulation in our previous work~\cite{Luna:2018jfk}, it is easy to check that the final BH ends up with $Q=0.97Q_{\rm max}$, well below the threshold for SCC violations.\footnote{The charge $Q$ of the BH, of course, does not change; what changes is $Q_{\rm max}$, since it depends on the BH's final mass.}
This was not recognized as a problem in the paper, since $Q_{\rm th}$ was mistakenly identified to be around $0.95Q_{\rm max}$. This erroneous value is also included in the paper's discussion. Why, then, were we not seeing mass inflation in all cases? The answer is that as the charge of the BH approaches $Q_{\rm th}$, so does the timescale for the onset of mass inflation (see Eq.~\eqref{vonLow}). Therefore, for a finite evolution time, one may be tempted to conclude in favor of SCC violation even for sub-critical spacetimes with $Q<Q_{\rm th}$, where mass inflation is expected to exist, especially if in that (short) evolution we observe an almost constant small Misner-Sharp mass accompanied by diverging curvature. That was exactly what happened with the evolutions that were erroneously associated with no mass inflation solutions in Ref.~\cite{Luna:2018jfk}; in fact, we now expect that in those evolutions mass inflation was ``just around the corner'', i.e., just after $\mathring{v}=50$, where unfortunately numerical noise becomes dominant and does not allow further conclusions.

In Fig.~\ref{Charged} we present new results concerning charged scalar fields (where numerical noise is smaller) by perturbing a scenario where the linear analysis also suggested a potential failure of SCC~\cite{Cardoso:2018nvb} (compare with~\cite{Mo:2018nnu,Dias:2018ufh}). In this case accretion again drives the black hole away from the region of expected SCC violation, and mass inflation does occur. Notice that the Misner-Sharp mass is almost constant well into the region where the curvature is already diverging.

\begin{figure}
\includegraphics[width = 0.49\textwidth]{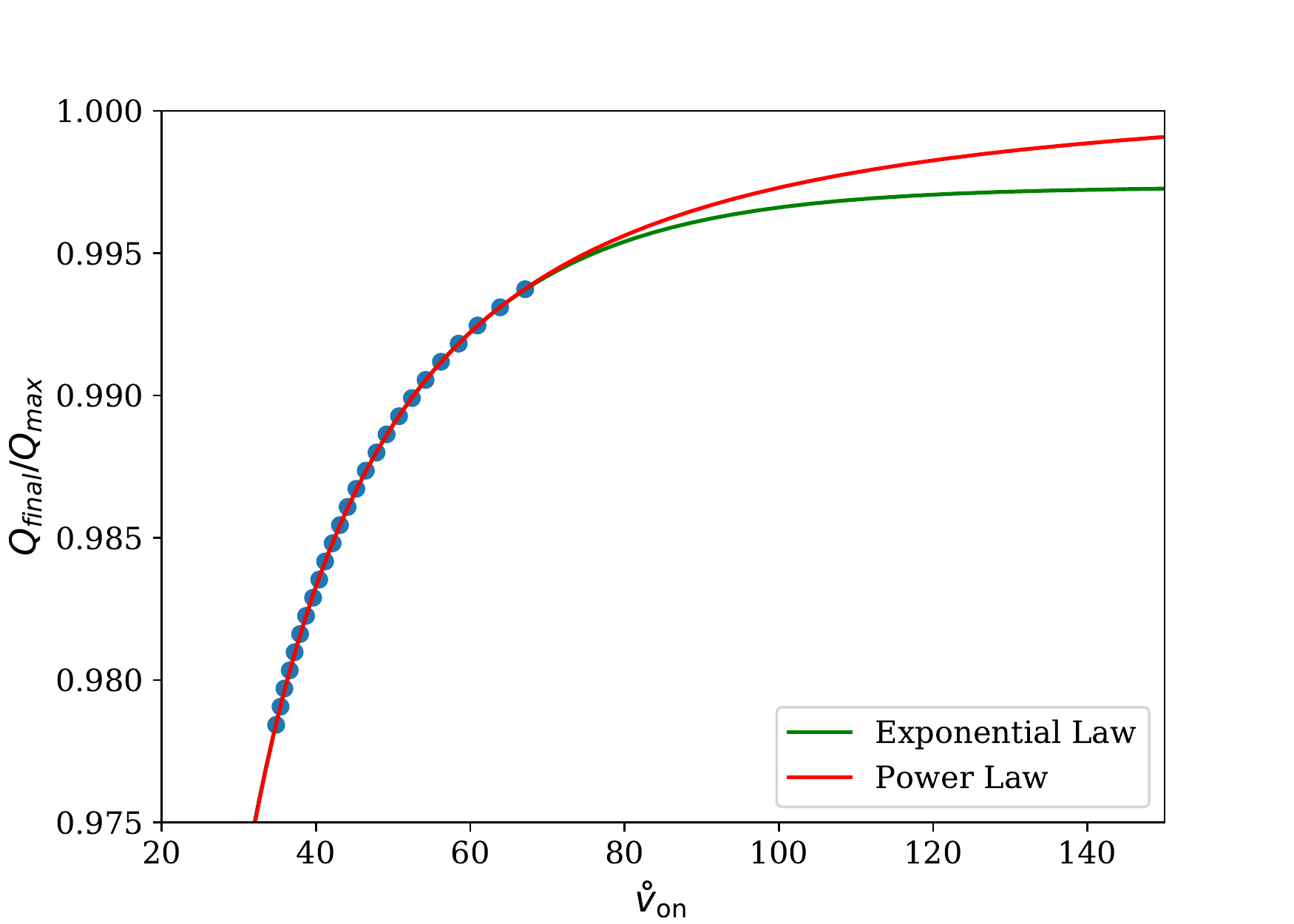}
\caption{Timescale for the onset of mass inflation -- given by the value $\mathring{v}=\mathring{v}_{\rm on}$ of the Eddington-Finkelstein coordinate where the Misner-Sharp mass increases above a threshold of $M=1.025~M_0$ -- as a function of the final charge. The scalar field charge is $q = 0.45$.
\label{Timescales}}
\end{figure}
It is quite clear that the onset of mass inflation (by which we mean the value $\mathring{v}=\mathring{v}_{\rm on}$ of the Eddington-Finkelstein coordinate along the outgoing null geodesics where the Misner-Sharp mass  increases above a threshold of $M=1.025~M_0$) is delayed as one gets closer and closer to maximally charged BHs. In fact, for the neutral case we can use the results in~\cite{Costa:2017tjc} to prove the lower bound
\begin{equation}
\label{vonLow}
\mathring{v}_{\rm on} (u) \geq C \log\left(\frac{1}{u-u_{\rm eh}}\right) \frac{1}{1-Q/Q_{\rm max}}\;,
\end{equation}
for all sufficiently small values of the ingoing null coordinate $u>u_{\rm eh}$, where $u=u_{\rm eh}$ is the event horizon and $C>0$ is a constant that depends on the final black hole parameters, but is bounded away from zero (see the Appendix for more details).

The question then reduces to determining the time required for the onset of mass inflation as a function of the final BH charge. More precisely, does this timescale become infinite before extremality? If it does, this would imply a violation of SCC, since it would leave a range of final charges where no mass inflation occurs; if it does not, SCC is preserved. Note that an estimate of the form~\eqref{vonLow} does not suffice to reach such a conclusion.

Figure~\ref{Timescales} depicts the (approximate) time for the onset of mass inflation as a function of the final charge. The different values of the final charge are achieved by performing simulations with different initial black hole charges $Q_0$. In this case we are using a charged scalar field with $q = 0.45$ (compare with Fig.~\ref{Charged} and Ref.~\cite{Cardoso:2018nvb}), and the code is only able to follow the evolution until $\mathring{v} \sim 70$.
We have fitted the numerical data with two different functions, a power-law $f(\mathring{v}_{\rm on}) = b_1 - \frac{k_1}{\mathring{v}_{\rm on}^{\alpha}}$ and an exponential $g(\mathring{v}_{\rm on}) = b_2 - k_2 e^{-\beta \mathring{v}_{\rm on}} - k_3 e^{-\gamma  \mathring{v}_{\rm on}^2}$. These two curves, even though they both agree very well with the numerical data, have very different asymptotic behaviors.
Quite obviously, then, the given data is not nearly enough to extrapolate the asymptotic behavior of the curve without knowing its specific form.

This takes us to the next problem: The smaller the perturbation, the longer it takes for the onset of mass inflation to occur. If the scalar field pulse is small enough to keep the BH in the right extremality range then we would need an extremely long evolution to rule out mass inflation. Such a long evolution seems to be well beyond the capabilities of our numerical scheme (and of any other double-null code that we have seen so far).

\begin{acknowledgments}

We thank H.~Zhang and Z.~Zhong for helpful discussions.
The work of R.L.\ is supported by funds from the Solvay family as well as by the F.R.S.-FNRS Belgium through conventions FRFC PDR T.1025.14 and IISN 4.4503.15.
M.Z.\ acknowledges financial support provided by FCT/Portugal through the IF
programme, grant IF/00729/2015.
J.L.C.\ and J.N.\ acknowledge financial support provided by FCT/Portugal through UIDP/MAT/04459/2020. 
The authors acknowledge financial support provided under the European Union's H2020 ERC 
Consolidator Grant ``Matter and strong-field gravity: New frontiers in Einstein's theory'' grant 
agreement no. MaGRaTh--646597. 
This project has received funding from the European Union's Horizon 2020 research and innovation programme under the Marie Sklodowska-Curie grant agreement No 101007855.
We acknowledge financial support provided by FCT/Portugal through grants PTDC/MAT-APL/30043/2017 and PTDC/FIS-AST/7002/2020.
The authors would like to acknowledge networking support by the GWverse COST Action CA16104, ``Black holes, gravitational waves and fundamental physics.''
\end{acknowledgments}


\appendix
\section{Derivation of Eq.~\eqref{vonLow}.}


In this appendix we provide the necessary details to derive Eq.~\eqref{vonLow} from the results in~\cite{Costa:2017tjc}.  
In that paper a curve $\gamma$, parameterized by $(u,v_{\gamma}(u))$, is constructed to probe the CH. It turns out, see the proof of Lemma 8.1, that in the past of that curve the Misner-Sharp mass (renormalized Hawking mass in the terminology of~\cite{Costa:2017tjc}) can be made arbitrarily close to the final black hole mass, by choosing $u-u_{\rm eh}>0$ sufficiently close to zero. From this we immediately see that  $\mathring{v}_{\rm on} (u) \geq v_{\gamma}(u)$.  It then follows, from the unnumbered equation following Eq. (171) of~\cite{Costa:2017tjc}, that 
\begin{equation}
\label{vonLow0}
\mathring{v}_{\rm on} (u) \geq v_{\gamma}(u) \geq C \log\left(\frac{1}{u-u_{\rm eh}}\right) \frac{1}{k_+}\;,
\end{equation}
where $k_+>0$ is the surface gravity of the event horizon of the reference RNdS solution with parameters prescribed by the final parameters of the dynamic solution under analysis. To derive the last estimate one needs to note that we are allowed to rescale the smallness parameter $\delta$ appearing in~\cite{Costa:2017tjc}  by $\delta k_+$ and that the constant $c>0$ of~\cite{Costa:2017tjc} approaches unity as $\delta k_+$ goes to zero.  

To finish the derivation of~\eqref{vonLow} we just need to recall the  following simple identities given in terms of the horizon radii of the reference RNdS solution:
\begin{equation}
k_+=\frac{\Lambda}{6}\frac{(r_c+2r_++r_-)(r_+-r_-)(r_c-r_+)}{r_+^2}\;,
\end{equation}
\begin{equation}
Q/Q_{\rm max}=\frac{r_-}{r_+}\;.
\end{equation}
\bibliography{references}

\end{document}